\documentclass[aps,pra,twocolumn,amsmath,amssymb,10pt]{revtex4-2}
\usepackage{amsmath,amssymb,graphicx,color,hyperref}
\usepackage{bm}
\usepackage{amsfonts}
\usepackage{bbold}
\newcommand{\A}{\mathbb{A}}

\newcommand{\C}{\mathbb{C}}

\newcommand{\fu}{\mathfrak{u}}

\newcommand{\fn}{{\,\mathfrak{n}\,}}

\newcommand{\fr}{\mathfrak{r}}

\def\CC{\mathbb{C}}

\newcommand{\bc}{\mathbf{c}}

\newcommand{\bt}{\mathbf{t}}

\newcommand{\bJ}{\mathbf{J}}
\newcommand{\bK}{\mathbf{K}}

\newcommand{\cH}{\mathcal{H}}

\newcommand{\cF}{\mathcal{F}}

\newcommand{\cM}{\mathcal{M}}

\newcommand{\cP}{\mathcal{P}}

\newcommand{\cR}{\mathcal{R}}

\newcommand{\cT}{\mathcal{T}}

\newcommand{\cW}{\mathcal{W}}

\newcommand{\be}{\begin{equation}}
	\newcommand{\ee}{\end{equation}}
\newcommand{\bea}{\begin{eqnarray}}
	\newcommand{\eea}{\end{eqnarray}}

\newcommand{\ed}{\end{document}}

\newcommand{\bi}{\begin{itemize}}
\newcommand{\ei}{\end{itemize}}

\newcommand{\bce}{\begin{center}}
\newcommand{\ece}{\end{center}}

\newcommand{\IM}{{\rm Im}}

\begin{document}

	\title{Transport effects in non-Hermitian nonreciprocal systems: General approach}
	\author{Hamed Ghaemi-Dizicheh
	}
	\email {hghaemidizicheh@ku.edu.tr}
\affiliation{$^1$Department of Physics and Astronomy, University of Texas Rio Grande Valley, Edinburg, Texas 78539, USA\looseness=-4
}
	\begin{abstract}
In this paper, we present a unifying analytical framework for identifying conditions for transport effects such as reflectionless and transparent transport, lasing, and coherent perfect absorption in non-Hermitian nonreciprocal systems using a generalized transfer matrix method. This provides a  universal approach to studying the transport of tight-binding platforms, including higher-dimensional models and those with an internal degree of freedom going beyond the previously studied case of one-dimensional chains with nearest-neighbor couplings. For a specific class of tight-binding models, the relevant transport conditions and their signatures of non-Hermitian, nonreciprocal, and topological behavior are analytically tractable from a general perspective. We investigate this class and illustrate our formalism in a paradigmatic ladder model where the system's parameters can be tuned to adjust the transport effect and topological phases.
	\end{abstract}
	\maketitle
\section{Introduction}
In recent years, the study of topological models has been extended to non-Hermitian Hamiltonians, which appear as effective Hamiltonians of open systems or wave systems with gain and loss \cite{moiseyev2011book,Ozawa2019topologicalphotonic,Yuto2020book}. The appearance of nonreciprocity introduces another non-Hermitian extension which gives rise to novel phenomena such as the non-Hermitian skin effect (NHSE) \cite{Yao2018skin,Yao2018skin2,Longi2019probingskin,schomerus2020nonreciprocal,Zhang2020skin,Okuma2020skin,Yi2020skin} and the recently identified edge burst effect \cite{Xue2022edgeburst}. In the former effect, the bulk states become localized at the edge of a finite system with open boundary conditions, a behavior that is drastically different from its periodic counterpart. In the latter effect, a substantial portion of loss occurs at the system boundary.\\
Besides, incorporating non-Hermitian topological models into photonics systems has introduced novel platforms known as topological photonics \cite{lu2014topological,Segev2021topologicalphotonicsrev,price2022roadmap}, which can induce a rich diversity of transport phenomena such as lasing \cite{Henning2010quantumnoise,Zhao2018microlasers,Parto2018ActiveArray,ghaemi2017nonlinearSS,ghaemi20202D}, coherent perfect absorption \cite{Longhi2010CPA,Chong2010CPA,Wang2021CPA}, reflectionless scattering \cite{Lin2011invisibility}, invisibility \cite{el2018non}, and transparency \cite{Longhi2015transparency}.\\ Hence, developing a formulation for the distinct transport effects in non-Hermitian nonreciprocal systems paves the way toward investigating interesting phenomena in such platforms. Following this motivation in our previous paper, we studied (i) conditions for variant transport effects, (ii) their compatibility with each other, and (iii) their adjustment by tuning suitable parameters facilitated by symmetry or topology \cite{ghaemi-dizicheh2021compatibility}. We then found distinct transport signatures of non-Hermitian, nonreciprocal, and topological systems. For instance, (I) the direction of reflectionless transport depends on the topological phases, (II) invisibility coincides with the non-Hermitian skin-effect phase transition of topological edge states, and (III) a unidirectional transparent coherent perfect absorption emerges.\\
Our previous study considered a generic one-dimensional chain with nearest-neighbor couplings. However, this study can be further generalized to more involved models. Here, we address this generalization to analytically formulate the spectral conditions for a range of distinct phenomena and identify their interdependence.\\
Our method is based on exploiting the transfer matrix \cite{Mos20} to characterize the transport boundary conditions. Along with the scattering matrix \cite{Sch13c}, the transfer matrix is a prominent mathematical instrument to study transport in finite-range scattering potentials. Recently, a developed version of the transfer matrix method has paved its way as an analytic approach in non-Hermitian tight-binding models to study topological \cite{dwivedi2016bulk,Kunst2019transfermatrix} and localization phenomena \cite{wang2022localisationtransfer}.\\ In Ref. \cite{ghaemi-dizicheh2021compatibility}, we employed the transfer matrix to define transport boundary conditions and their connection. Following the same formulation, we can extend the derivation of the transfer matrix describing transportation in a general nonreciprocal lattice. This approach can provide a universal characterization of various transport phenomena.\\ We then consider a class of nonreciprocal tight-binding systems whose transport signatures are described by a $2\times 2$ transfer matrix. We will show that our conclusions in \cite{ghaemi-dizicheh2021compatibility} for the transport effects and their compatibility can be retrieved in a general context for all those models settled in this specific category.\\
The rest of this paper is organized as follows: we describe our method for constructing the transfer matrix for a general tight-binding model in real and propagating space in section \ref{sectionII}, where we present a global scheme of the leads in Fig.\ref{fig1}. We then determine the boundary conditions of distinct transport effects and their interplay in section \ref{section3} with a general attitude.\\
In section \ref{section4}, we investigate our general formalism for a class of systems whose transfer matrix reduces to a $2 \times 2$ matrix which is familiar in the photonics models. For these models, we can categorize the transport boundary conditions as implicit equations in terms of the parameters, possibly associated with the topology and reciprocity of the system. To demonstrate our approach, we investigate the transport effect of the Ladder lattice. We then show that by generalizing the coupling parameters, the system can be tuned to be reflectionless in the non-trivial topological phase. This suggests that our research can be used to determine an appropriate set of system parameters for building a platform with specific transport effects in different topological phases.
	\section{Mathematical Set-up.}\label{sectionII}
\subsection{Model of Supercells}
\begin{figure}[t]
	\includegraphics[scale=0.32]{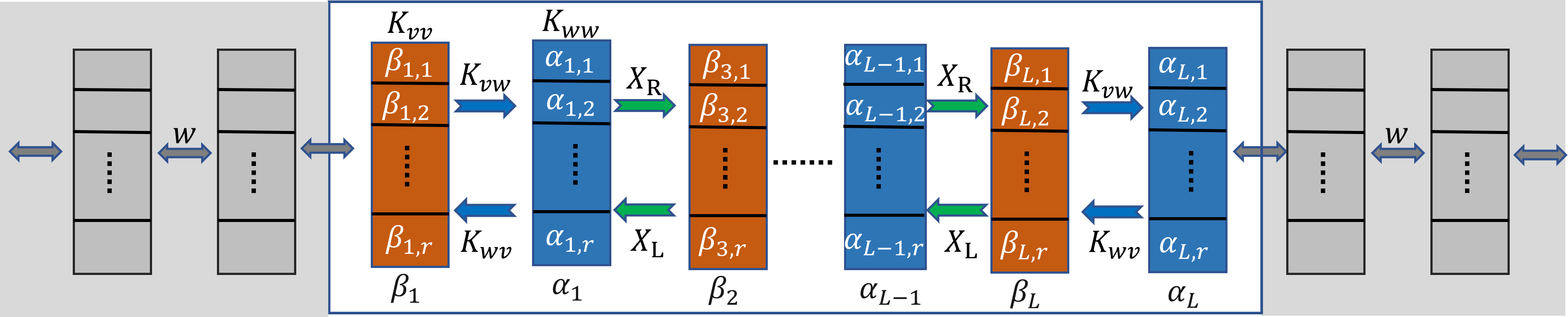}
	\caption{A schematic diagram of the recursion relations (\ref{rec1}) and (\ref{rec2}) describing non-Hermitian nonreciprocal systems with $q$ internal degrees of freedom and range of interaction $\fr$. This diagram represents a general tight-biding model connecting to the featureless leads (with coupling $w<0$).} \label{fig1}
\end{figure}

To extend the theory of transport from a rather simple one-dimensional chain to a more complicated structure, we consider a system that is described by a tight-binding model whose Hamiltonian is given by
\bea
&&\cH=\sum_{n=1}^{N}\sum_{\alpha,\beta=1}^{q}\left[\sum_{l=1}^{\fr}(c^{\dagger}_{n,\alpha}[\bt^{L}_{l}]_{\alpha\beta}c_{n+l,\beta}+c^{\dagger}_{n+l,\alpha}[\bt^{R}_{l}]_{\alpha\beta}c_{n,\beta}\right.  \nonumber\\
&&\left. ~~~~~~~~~~~~~~~~~~~+c^{\dagger}_{n,\alpha}[\bt_{0}]_{\alpha\beta}c_{n,\beta}) \right],\label{Hamiltonian of supercell}
\eea
where $ \bt^{L(R)}_{l} $ denotes the hopping matrix to the left (right), and $ \bt_{0} $ is the intra-unit-cell term. Also, in the Hamiltonian (\ref{Hamiltonian of supercell}), the parameter $ \fr $ is the range of hopping, and $ q $ determines the number of internal degrees of freedom, e.g., spin, orbital, or sublattice, per unit-cell.\\ Following the main idea developed in \cite{dwivedi2016bulk,Kunst2019transfermatrix}, for the construction of the transfer matrix, we consider a bundle of $ \fr $ adjacent sites, which reduces the model with $ N $ cells to a supermodel with $ L$ supercells, each containing $ \fn=q\fr $ degrees of freedom. Thus, the Hamiltonian of the superchain is given by
\be
H=\sum_{n=1}^{2L}[\bc^{\dagger}_{n}\bJ_{L}\bc_{n+1}+\bc^{\dagger}_{n}\bK\bc_{n}+\bc_{n+1}^{\dagger}\bJ_{R}\bc_{n}].
\ee
Here, we introduce $ \bJ_L $ ($ \bJ_R $) and $ \bK $, respectively, as the corresponding left (right) hopping and on-site matrices. The single-particle Schr\"{o}dinger equation ($ H\Psi =E\Psi$) is reduced to the following recursion relation
\be
E\Psi_{n}=\bK\Psi_{n}+\bJ_{R}\Psi_{n-1}+\bJ_{L}\Psi_{n+1},\label{recursion rel.}
\ee
with $ \Psi_{n}=\left(\psi_{2n-1},\cdots,\psi_{2n+\fr-1} \right)^{T} \in\CC^{\fn}$ defining wave function for each supercell. In the case of reciprocal transport, one has $ \bJ_{R}=\bJ^{\dagger}_{L} $, and the system is governed by a hermitian Hamiltonian with the real-valued spectrum provided that $\bK=\bK^{\dagger}$.\\
One can reduce the relation (\ref{recursion rel.}) by applying the reduced singular value decomposition (SVD) method \cite{strang2009introduction} on both $ \bJ_{L} $ and $ \bJ_{R} $. This reduction results in
\begin{align}
	&\bJ_{L}=VX_{L}W^{\dagger},&&\bJ_{R}=WX_{R}V^{\dagger},\label{SVD}
\end{align}
where $ X_{L/R}=\rm diag\lbrace \xi^{1}_{L/R},\cdots\xi^{r}_{L/R}\rbrace  $ is a diagonal matrix of singular values ($ \xi^{1,2,\cdots,r}_{L/R} $) with $ r:=\rm rank(\bJ_{L}) $, which is defined by the number of its independent rows. The columns of $ V $ and $ W $ are called the left- and right-singular vectors of $ \bJ_L $\cite{Note1}. We further demand the size of the supercells to be big enough such that $ \bJ_{L/R} $ becomes nilpotent of degree 2, i.e., $ \bJ_{L/R}^{2}=0 $. It then follows $ r\leq q\fr $ \cite{dwivedi2016bulk,Kunst2019transfermatrix}. This condition leads to the following relations for left- and right-singular vectors
\begin{align}
	&V^{\dagger}V=W^{\dagger}W=\mathbb{1}_r,&&V^{\dagger}W=0,\label{orthonormal relation}
\end{align}
where $ \mathbb{1}_r $ is an $ r \times r $ identity matrix. According to the above relation span $ \lbrace V\rbrace $ and span $ \lbrace W\rbrace $ provide orthonormal basis of $\CC^{\fn} $ such that $\Psi_{n}$ can be expanded in the following form
\be
\Psi_{n}=V\alpha_{n}+W\beta_{n}+Y\zeta_{n},\label{expand superstate}
\ee
with $ Y $ is defined analogous to $ V $ and $ W $
\begin{align}
	&V^{\dagger}\cdot Y=W^{\dagger}\cdot Y=0,&&Y^{\dagger}\cdot Y=\mathbb{1}_r,\label{extended orthonormal relation}
\end{align}
and coefficients $ \alpha $, $ \beta $ and $ \zeta $ are
\bea
&&\alpha_{n}=(\alpha_{n,1},\alpha_{n,2},\cdots\alpha_{n,r}),\nonumber\\
&&\beta_{n}=(\beta_{n,1},\beta_{n,2},\cdots\beta_{n,r}),\nonumber\\
&&\zeta_{n}=(\zeta_{n,1},\zeta_{n,2},\cdots\zeta_{n,r}).
\eea
By substituting hopping matrices ($ J_{L,R} $) and superstate ($ \Psi_{n} $) respectively from Eqs. (\ref{SVD}) and (\ref{expand superstate}), the right side of recursion relation (\ref{recursion rel.}) reduces to
\be
E\Psi_{n}=\bK[V\alpha_{n}+W\beta_{n}]+WX_{R}\alpha_{n-1}+VX_{L}\beta_{n+1}.
\ee
Here, we assume that the only relevant direction here is span $ \lbrace V\rbrace $ and span $ \lbrace W\rbrace $. The relation above can be reduced further by multiplying $ W^{\dagger} $ and $ V^{\dagger} $ from the left and making use of orthonormal relations (\ref{orthonormal relation}). Then, we obtain
\bea
&E\mathbb{1}_{r}\beta_n=K_{ww}\beta_n+X_{R}\alpha_{n-1}+K_{wv}\alpha_n,&\label{rec1}\\
&E\mathbb{1}_{r}\alpha_n=K_{vv}\alpha_n+X_{L}\beta_{n+1}+K_{vw}\beta_n,\label{rec2}&
\eea
where $ K_{ab}:=A^{\dagger}\bK B\in \text{Mat}(r,\CC) $ with $ A,B\in \lbrace V,W\rbrace $. In this formulation, the superlattice is restricted to the region $ 0\leq n\leq 2L$, while the remaining sites represent the lead structure. Similar to the approach presented in \cite{ghaemi-dizicheh2021compatibility}, we model the lead's structure in the featureless wide-band limit which is obtained by setting $ K_{vw}=K_{wv}=X_{R}=X_{L}=w\mathbb{1}_{r} $ with $ w<0 $ for $ n\leq0 $ (left lead), and $ n\geq 2L $ (right lead). Besides, for each supersite in the lead, we tune the potential energy to the band center (i.e., $ K_{ww}=K_{vv}=E\mathbb{1}_{r} $). Then, we can introduce the propagating modes, such as
\bea
&&\beta_n=\phi^{(+)}i^{n-n_{+}^{\beta}},~~~~~~\alpha_n=\phi^{(+)}i^{n-n_{+}^{\alpha}},~~~~~~~\text{(right)}\nonumber\\
&&\beta_n=\phi^{(-)}(-i)^{n-n_{-}^{\beta}},~\alpha_n=\phi^{(-)}(-i)^{n-n_{-}^{\alpha}},~~~\text{(left)}\nonumber\\\label{propagating mode}
\eea
where the amplitude $ \phi^{(\pm)}=(\phi^{(\pm)}_1,\cdots,\phi^{(\pm)}_r) $ are position independent scattering modes throughout the lead's structure. The possible non-integer offsets $ n_{\pm}^{\alpha,\beta} $ can be chosen separately in each lead and account for the $U(1)$ gauge freedom. We schematically demonstrate the lead's structure in diagram \ref{fig1}.
\subsection{Transport Framework}
As the main step to characterize the transport properties of the system, we construct the transfer matrix in the space of real- and propagating-state. By introducing $ \Phi_{n}=(\mathbb{\beta}_{n},\mathbb{\alpha}_{n-1})^{T} $, we can define the transfer matrix $ M $ that connects the states in adjacent supercell through the following relation
\be
\Phi_{n+1}=M_n\Phi_{n}.\label{definition of transfer matrix}
\ee
In light of recursion relations (\ref{rec1}) and (\ref{rec2}), one can find the explicit form of the real-space transfer matrix as
\begin{equation}
	M_n=\left( \begin{array}{cc}
		X_{L}^{-1}[G_{vv}K_{wv}^{-1}G_{ww}-K_{vw}]&-X_{L}^{-1}G_{vv}K_{wv}^{-1}X_{R}  \\
		\\
		K_{wv}^{-1}G_{ww}&-K_{wv}^{-1}X_{R}
	\end{array}\right),\label{Transfer matrix of supercells}
\end{equation}
where $ G_{ab}=A^{\dagger}(E\mathbb{1}_{r}-\bK)B\in \text{Mat}(r,\CC) $ with $ A,B\in \lbrace V,W\rbrace $. 
The size of the transfer matrix depends on the rank of the hopping matrix $ \bJ_{L} $. Then, the $2r\times2r$ transfer matrix for the whole system can be given as
\be
M=M_L\cdots M_3\cdot M_2\cdot M_1.\label{real-space transfer matrix}
\ee
By considering the propagating mode (\ref{propagating mode}) of the superlattice, the transfer matrix in the propagating-state basis is then defined by
\bea
\left(\begin{array}{c}
	\phi^{(+,R)}    \\
	\phi^{(-,R)}  
\end{array}\right)=\cM\left(\begin{array}{c}
	\phi^{(+,L)}    \\
	\phi^{(-,L)}   
\end{array}\right),\label{def.propagating-space transfer matrix}
\eea
where $ L $ and $ R $ refer to the left and right lead. This transfer matrix captures the transport features in terms of linear relations between the propagating wave amplitudes in the leads. It is given by invoking the wave-matching condition of propagating waves. We then obtain
\be
\cM=\frac{1}{2}\left(\begin{array}{cc}
	-i\mathbb{1}_r&\mathbb{1}_r  \\
	\\
	\mathbb{1}_r&-i\mathbb{1}_r
\end{array}\right)M\left(\begin{array}{cc}
	i\mathbb{1}_r&\mathbb{1}_r  \\
	\\
	\mathbb{1}_r&i\mathbb{1}_r
\end{array}\right).\label{scattering-space transfer matrix}
\ee
In the next step, we can find the $ r \times r $ right and left reflection and transmission matrices in terms of the entries of the transfer matrix $ \cM $ in the propagating-state basis as
\begin{align}
	&\left( \begin{array}{cc}
		\textbf{r}	& \textbf{t}' \\
		\textbf{t}	& \textbf{r}'
	\end{array}\right)=&\nonumber\\
	&\left(\begin{array}{cc}
		-\cM_{22}^{-1}\cdot\cM_{21}&\dfrac{\widetilde{\cM}_{22}-\widetilde{\cM}_{21}\cdot\widetilde{\cM}_{11}^{-1}\cdot\widetilde{\cM}_{12}}{\vert \cM \vert}  \\
		\\
		\cM_{11}-\cM_{21}\cdot\cM_{22}^{-1}\cdot\cM_{21}&-\dfrac{\widetilde{\cM}_{11}^{-1}\cdot\widetilde{\cM}_{12}}{\vert \cM \vert}
	\end{array}\right),&\nonumber\\
\end{align}
where $ \widetilde{\cM}_{ij} $ are the enteries of the adjugate matrix of $ \cM $ and $ \vert\cM\vert $ is the determinant. 
\section{Boundary conditions: General Approach}\label{section3}
The transfer matrix approach in real-space ($M$) and scattering-state ($\cM$) basis provides us with a comprehensive algorithm describing different boundary conditions for a variety of physical effects for the supermodel depicted in Fig. \ref{fig1}. This section defines various physical transport phenomena in the supermodel and demonstrates how we can characterize them using the transfer matrix method.
\subsection{Periodic boundary conditions}
The periodic boundary conditions for the system with $ L $ supercells demands $ \Psi_{n+N}=e^{ik_x}\Psi_n $. By considering decomposition of $ \Psi $ (\ref{expand superstate}) and the definition of the transfer matrix (\ref{definition of transfer matrix}), it leads to
\be
M(E,k_\bot)\Phi_n=e^{ik_x}\Phi_n. \label{periodic}
\ee
Here, we consider a system in $d$ spatial dimensions with open boundary conditions along $x$-coordinate and periodic boundary conditions along the remaining $d-1$ directions, which are parametrized by the transverse quasimomentum $k_{\bot}\in \cT^{d-1}$. This system can also be interpreted as a family of one-dimensional chains parametrized by $k_{\bot}$. Eq. (\ref{periodic}) reveals that the system with periodic boundary conditions has a state with complex energy $ E $ if and only if $ e^{ik_x}\in \text{Spec}[M(E,k_\bot)] $. The complex spectrum of the system forms a closed curve in the complex plane, satisfying the following implicit dispersion equation
\be
\det(M(E,k_\bot)-e^{ik_x}\mathbb{1}_r)=0.
\ee 
In the propagating-wave space, by considering the unitary transformation (\ref{scattering-space transfer matrix}), the form of implicit dispersion equation turns into
\be
\det(\cM(E,k_\bot)-e^{ik_x}\mathbb{1}_{r})=0.
\ee 
\subsection{Open boundary conditions}
For the truncated system, the open boundary conditions can be fulfilled by setting $ \Psi_0=\Psi_{N+1}=0 $, which leads to $ \alpha_0=\beta_{L+1}=0 $ at the first sites of the leads. In terms of the real-space transfer matrix, the open boundary conditions imply
\bea
\left(\begin{array}{c}
	0    \\
	\alpha_{N}  
\end{array}\right)=M(E,k_\bot)\left(\begin{array}{c}
	\beta_{1}    \\
	0   
\end{array}\right)\textcolor{blue}{,}\label{OBC}
\eea
that corresponds to
\be
M_{11}=\mathbb{0}_r,
\ee 
where $ \mathbb{0}_r $ is a $ r\times r $ null matrix. The above relation can be read as $ r^2 $ separate implicit equations for the bound-state energies $ E_n $, whose solutions are generally discrete but possibly complex.
\subsection{Quasi-Bound state and lasing}
We define a quasi-bound state as a solution in which there only exists a purely outgoing wave (incoming waves are absent). This state can be achieved by setting $ \phi^{(+,L)}=0 $ into the left lead and $ \phi^{(-,R)}=0 $ into the right lead. In light of definition (\ref{def.propagating-space transfer matrix}), to achieve a quasi-bound state, $\cM_{22}$ component needs to fulfill
\be
\cM_{22}(E)=\mathbb{0}_r.\label{Quasi BCs}
\ee
Generally, those solutions that satisfy the quasi-bound boundary conditions are discrete and complex. In a special case, these solutions present a stationary lasing state when their imaginary part is zero, i.e., $ \IM(E_{n})=0 $.
\subsection{Coherent Perfect Absorption}
For the coherent perfect absorption (CPA), there is only a purely incoming wave (this corresponds to the time reversal of stationary lasing). Following the footsteps of the quasi-bound state boundary conditions, the CPA is achieved by setting $ \phi^{(-,L)}=0 $ into the left or ($ \phi^{(+,R)}=0 $) into the right lead, which conduces to
\be
\cM_{11}(E)=\mathbb{0}_r.\label{CPA BCs}
\ee 
\subsection{Reflectionless, Transparency, and Invisibility}
We define a reflectionless system from the left or right if the reflection matrix satisfies
\be
\textbf{r}(E)=\mathbb{0}_r~~~~~\text{or}~~~~~\textbf{r}'(E)=\mathbb{0}_r.
\ee
In terms of the transfer matrix, it can be expressed as
\be
\cM_{22}^{-1}\cdot\cM_{21}=\mathbb{0}_r~~~~~\text{or}~~~~~\widetilde{\cM}_{11}^{-1}\cdot\widetilde{\cM}_{12}=\mathbb{0}_r.
\ee
For the transparency condition, we consider just a phase shift for a probing wave passing through the system. In the language of the transmission matrix, depending on the side from which the system is probed, we have
\be
\textbf{t}(E)=\mathbb{0}_r~~~~~\text{or}~~~~~\textbf{t}'(E)=\mathbb{0}_r,
\ee
where in the context of the transfer matrix, it leads to
\bea
&\cM_{11}-\cM_{21}\cdot\cM_{22}^{-1}\cdot\cM_{21}=(-i)^{N}\mathbb{1}_r~~\text{or}&\nonumber\\
\nonumber\\
&\dfrac{\widetilde{\cM}_{22}-\widetilde{\cM}_{21}\cdot\widetilde{\cM}_{11}^{-1}\cdot\widetilde{\cM}_{12}}{\vert \cM \vert}=(-i)^{N}\mathbb{1}_r.&
\eea
The system is invisible to the left (right) source if it is simultaneously reflectionless and transparent under illumination from that side. To be invisible to the detector, one needs to make the system reflectionless under illumination from that side and transparent under exposure from the other side.
\section{Special Case: $ r=1 $}\label{section4}
\subsection{Classification of transport effects}
In this section, we focus on a specific case in which the rank of each hopping matrices $\bJ_{L, R}$ is one (i.e., $ r=1 $), representing a family of tight-binding models, the most famous of which are two-band lattices, such as the Su-Schrieffer-Heeger (SSH) model, are the famous ones. Other examples are the Chern insulator \cite{kohei2018chern}, 2D Dirac semimetal, and Hofstadter model \cite{Chernodub2015hofstadter}. In such a special family, various qualitative topological properties of these systems can be extracted from the $ 2 \times 2 $ transfer matrix \cite{dwivedi2016bulk,Kunst2019transfermatrix}. Our approach can be extended to $r
	\neq 1$ models, but presenting an analytic framework will be challenging. In the appendix, we derive a $4 \times 4$ transfer matrix for one of these models.\\ For $r=1$, the entries of the transfer matrix are given by complex-valued numbers, where the determinant is
\be
d=\dfrac{K_{vw}X_R}{K_{wv}X_L}.
\ee
From now on, $ K_{ab} $, $ G_{ab} $, and $ X_{L,R} $ represent complex-valued numbers instead of matrices. For the Hermitian case (i.e., $\bK^{\dagger}(E^{*})=\bK(E)$ and $X_R=X_L$), one can show that the determinant transforms to $d=K_{vw}(E)/K^{*}_{vw}(E^{*})$. Therefore, in the Hermitian system, for the real energies, the determinant lies on the unit circle since $d=\text{exp}[2i\text{arg}K_{vw}(E)]$. One can reach the same result when the system is non-Hermitian, but couplings and on-site matrices are invariant under the parity ($P$) and time-reversal ($T$) operator. Then the determinant lies on the unit circle in the $PT$-unbroken phase. If we consider the system to be periodic, then by setting the unit-cell transfer matrix of the superlattice to $ M_c $, the transfer matrix of the whole system with $ L $ supercell can be written as
\be
M(E,k_\bot)=d^{L/2}\left[U_{L-1}(z)\overline{M}_c-U_{L-2}\mathbb{1}_{2\times2} \right] ,
\ee
where $ \overline{M}_c=M_{c}/\sqrt{d} $, $ U_L(z) $ are the Chebyshev polynomials of the second kind and 
\be
z=\text{tr}\overline{M}_c/2=\dfrac{t}{2\sqrt{d}}=\dfrac{G_{vv}G_{ww}-K_{vw}K_{wv}-X_{L}X_{R}}{2\sqrt{K_{wv}K_{vw}X_{L}X_{R}}}.\label{z-parameter}
\ee
As a first consequence of the general formalism we introduced in this paper, one can formulate a set of boundary conditions for the particular case $ r=1 $. In table (\ref{table1}), we characterize the transport properties of the system in terms of the entries of the transfer matrix and give the corresponding implicit equations.\\
In stationary transport phenomena such as reflectionless, transparency, and invisibility, real-valued energy is given. Hence the corresponding implicit equations in the table (\ref{table1}) can define the possible range of couplings or on-site potential for the given energy (or frequency) in which the system is in a stationary transport state. For other boundary conditions, the solutions of the corresponding implicit equations lead to a discrete complex spectrum describing quasi-stationary behavior.\\ There exists an alternative approach to obtain the spectrum related to open, quasi-band, and scattering zero boundary conditions. Consider eigenstates and eigenenergies of the transfer matrix satisfying
\be
M_c(E)\varphi_{1,2}=\rho_{1,2}\varphi_{1,2}.
\ee
In the case of diagonalizable $ M_c $, $ \varphi_{1,2}$ form a basis of $ \C^2 $ in which one can expand $ \Phi_{N} $, and the coefficients can be defined by considering stationary boundary conditions \cite{Note2}. For instance, the open boundary conditions (\ref{OBC}) can be translated to
\bea
&\left(\begin{array}{c}
	\beta_{1}    \\
	0  
\end{array}\right)=a_1\varphi_1+a_2\varphi_2,&\left(\begin{array}{c}
	0   \\
	\alpha_{N}  
\end{array}\right)=a_1\rho_1^N\varphi_1+a_2\rho_2^N\varphi_2.\nonumber\\
\eea
These equations reduce to a set of two complex homogeneous linear equations in terms of two variables $ a_1 $ and $ a_2 $, satisfying
\be
\sum_{l=1}^2 a_l\cP_+\varphi_l=\sum_{l=1}^2 a_l\rho_l^N\cP_-\varphi_l=0,\label{OBC Matrix Eq.}
\ee
where $ \cP_{\pm}:\CC^2\rightarrow\CC $ are projection operators which are defined as $ \cP_+=(0,1) $ and $ \cP_-=(1,0) $. They inject the state $\Phi$ into the subspaces $\beta$ and $\alpha$, respectively. Indeed, the open boundary condition \ref{OBC Matrix Eq.} can be interpreted as the Dirichlet condition on the left edge, which is equivalent to the statement that $\Phi$ belongs to the range of $\cP_+$, while the right edge is equivalent to the statement that the state $\Phi$ belong to the range of $\cP_-$ \cite{dwivedi2016bulk}.\\ One can recast (\ref{OBC Matrix Eq.}) in a matrix equation, which, according to Cramer's rule, has a nontrivial solution if
\be
\text{det}(\cR_1^N\varphi_1~\cR_2^N\varphi_2)=0,\label{determinant OBC}
\ee
where
\be
\cR_l=\left( \begin{array}{cc}
	\rho_{l}	&0  \\
	0& 1
\end{array}\right).
\ee
Equation (\ref{determinant OBC}) can be solved to get the set of energies for which the system with open boundary conditions has eigenstates.\\
Following the same approach for the quasi-bound and CPA boundary conditions, we reach to
\be
\sum_{l=1}^2 a_l\cP_{\mp}\tilde{\varphi}_l=\sum_{l=1}^2 a_l\rho_l^N\cP_{\pm}\tilde{\varphi}_l=0,\label{Quasi and CPA Matrix Eq.}
\ee
where the negative (positive) sign in each term refers to quasi-bound (CPA) boundary conditions and $ \tilde{\varphi}_l $ are the eigenstates of the transfer matrix in propagating state, $\cM$. They are given by
\be
\tilde{\varphi}_l=\frac{1}{\sqrt{2}}\left( \begin{array}{cc}
	i	&1  \\
	1& i
\end{array}\right)\varphi_l.
\ee
The corresponding energies can be obtained by finding the solutions of
\be
\text{det}(\cR_1^N\tilde{\varphi}_1~\cR_2^N\tilde{\varphi}_2)=0,
\ee
for the CPA and
\be
\text{det}(\tilde{\cR}_1^N\tilde{\varphi}_1~\tilde{\cR}_2^N\tilde{\varphi}_2)=0,
\ee
for the quasi-bound, where
\be
\tilde{\cR}_l=\left( \begin{array}{cc}
	1	&0  \\
	0& \rho_{l}
\end{array}\right).
\ee
\begin{table*}[th]
	\begin{tabular}{ccc}
		\hline
		Boundary Conditions & Characterization (arbitrary $ r $) & Implicit Equation ($ r=1 $) \\
		\hline\\
		open & $
		M_{11}=\mathbb{0}_r $ & $  \sqrt{\dfrac{d}{\kappa}}\left[\dfrac{G_{vv}G_{ww}}{w^{2}} -\kappa\right] =\dfrac{U_{L-2}(z)}{U_{L-1}(z)}
		$
		
		\\[.8cm]
		periodic & $ \det(M-e^{ik_x}\mathbb{1}_{2r \times 2r})=0 $ & $t-d e^{-ik_x}-e^{ik_x}=0$
		
		\\[.8cm]
		quasi-bound & $ \cM_{22}=\mathbb{0}_r $ & $U_{L-1}(z)\sqrt{\dfrac{d}{\kappa}}\left[(G_{vv}-iw)(G_{ww}-iw)-w^{2}\kappa \right] -2w^{2}U_{L-2}(z)=0$
		\\[.8cm]
		CPA & $ \cM_{11}=\mathbb{0}_r $ & $U_{L-1}(z)\sqrt{\dfrac{d}{\kappa}}\left[(G_{vv}+iw)(G_{ww}+iw)-w^{2}\kappa \right]-2w^{2}U_{L-2}(z)=0$
		\\[.8cm]
		right-reflectionless & $ \widetilde{\cM}_{11}^{-1}\cdot\widetilde{\cM}_{12}=\mathbb{0}_r $ & $ U_{L-1}(z)\left[(G_{ww}-iw)(G_{vv}+iw)-w^{2}\kappa \right] =0
		$
		\\[.8cm]
		left-Reflectionless & $ \cM_{22}^{-1}\cdot\cM_{21}=\mathbb{0}_r $ & $U_{L-1}(z)\left[(G_{ww}+iw)(G_{vv}-iw)-w^{2}\kappa \right] =0$
		\\[.8cm]
		right-invisible & $ 
		\begin{cases} \widetilde{\cM}_{11}^{-1}\cdot\widetilde{\cM}_{12}=\mathbb{0}_r,\\\cM_{11}-\cM_{21}\cdot\cM_{22}^{-1}\cdot\cM_{21}=(-i)^{N}\mathbb{1}_r \end{cases} $ & $ 
		\begin{cases} \sqrt{\dfrac{d}{\kappa}}\left[(G_{vv}-iw)(G_{ww}-iw)-w^{2}\kappa \right]=\dfrac{(-i)^{N}d-2w^{2}U_{L-2}(z)}{U_{L-1}(z)}  ,\\U_{L-2}(z)=-(-i)^{N}d \end{cases} $
		\\[.8cm]
		left-invisible & $ 
		\begin{cases} \cM_{22}^{-1}\cdot\cM_{21}=\mathbb{0}_r ,\\\dfrac{\widetilde{\cM}_{22}-\widetilde{\cM}_{21}\cdot\widetilde{\cM}_{11}^{-1}\cdot\widetilde{\cM}_{12}}{\vert \cM \vert}=(-i)^{N}\mathbb{1}_r \end{cases} $ & $ 
		\begin{cases} \sqrt{\dfrac{d}{\kappa}}\left[(G_{vv}-iw)(G_{ww}-iw)-w^{2}\kappa \right]=\dfrac{(-i)^{N}-2w^{2}U_{L-2}(z)}{U_{L-1}(z)}  ,\\U_{L-2}(z)=-(-i)^{N}   \end{cases} $
		\\[.8cm]
		\hline
	\end{tabular}
	\caption{The second row of the table compares distinct boundary conditions (BCs) for a general model illustrated in Fig.\ref{fig1} where the transfer matrix has to be taken as functions of energy $E$. For those models with $r=1$, the third row of the table characterizes implicit equations in terms of energy $E$ for transport effects listed in the first row. The stationary situations correspond to given real energy, while those equations result in complex energies refer to quasi-stationary effects.}\label{table1}
\end{table*}
\subsection{Compatibility of transport effects}
In this part, we take a look at some of the aspects driven by the corresponding boundary conditions in the context of the transfer matrix. In addition to the determinant $d$, we introduce the followng parameter
\begin{align}
\kappa=\dfrac{K_{vw}K_{wv}}{w^2}.
\end{align}
As shown in Ref. \cite{ghaemi-dizicheh2021compatibility}, the two parameters $d$ and $\kappa$ play an important role in specifying the reciprocity and topology signature in the one-dimensional SSH model. Indeed, $d$ quantifies the amount of nonreciprocity where the system becomes reciprocal and the non-Hermitian skin effect disappears for the parameter in which the transfer matrix is unimodular. The parameter $\kappa$ represents the topological characteristic, with $\kappa=1$ corresponding to the topological phase transition.\\ The transfer matrix (\ref{Transfer matrix of supercells}) of the unit-cell can be written in therms of the these parameters such as
\begin{equation}
	M_c=\sqrt{\frac{d}{\kappa}}\left( \begin{array}{cc}
		\dfrac{G_{vv}G_{ww}}{w^2}-\kappa&-\dfrac{G_{vv}}{w}  \\
		\\
		\dfrac{G_{ww}}{w}&-1
	\end{array}\right).\label{Transfer matrix of supercells for r1}
\end{equation}
Following the corresponding implicit equations from the table \ref{table1} for the right and left reflectionless transport yields
\be
\cF U_{L-1}(z)=0,\label{reflectionless BC}
\ee
with
\be
\cF=G_{vv}G_{ww}+w^2(1-\kappa)\mp iw(G_{ww}-G_{vv}),\label{reflectionless for r1}
\ee
where the minus (plus) sign applies to probing the system from the left (right). Condition (\ref{reflectionless BC}) represents two different types of solutions where we interpret ($ U_{L-1}(z)=0 $) as a global and ($ \cF=0 $) as a local mechanism. The solutions of the global mechanism given by the Chebyshev nodes depend on the length of the system $L$. If we set $ E=K_{vv} $ or $ E=K_{ww} $ which corresponds to $ G_{vv}=0 $ or $ G_{ww}=0 $ respectively \cite{Note3}, then from Eq. (\ref{z-parameter}), one can find
\be
z=-\frac{1}{2}\left( \sqrt{\kappa}+\frac{1}{\sqrt{\kappa}}\right) .
\ee
It is easy to check that $ z $ remains intact under the transformation $ \kappa\rightarrow 1/\kappa $ showing that the global mechanism for the reflectionless boundary conditions is also invariant under this transformation. \\ This is in contrast with the local mechanism. To reveal the difference between global and local mechanisms, consider probing the system from the left (right). Then the local condition leads to
\be
w(1-\kappa)=\pm i(K_{ww}-K_{vv}),\label{general reflectionless}
\ee
where the plus (minus) sign gives left (right)-reflectionless. According to \ref{general reflectionless}, in the local mechanism, $ \kappa $ shows up as an essential parameter for both left and right reflectionless boundary conditions.\\
To study other transport effects, we express the transfer matrix in the following form
\begin{equation}
	\tilde{\cM}=\left( \begin{array}{cc}
		\kappa_{+}+\dfrac{i\kappa_{-}\tilde{K}}{w(\kappa-1)}&-i\kappa_{-}\left[1-\dfrac{i\tilde{K}}{w(\kappa-1)}\right]  \\
		\\
		i\kappa_{-}\left[ 1+\dfrac{\tilde{K}}{w(\kappa-1)}\right]&\kappa_{+}-\dfrac{i\kappa_{-}\tilde{K}}{w(\kappa-1)}
	\end{array}\right),\label{}
\end{equation}
where $\kappa_{\pm}:=(\kappa^{L}\pm 1)$, $\tilde{K}:=K_{ww}-K_{vv}$ and also we consider normalized transfer matrix $\tilde{\cM}:=\frac{2\kappa^{L/2}}{(-1)^{L}d^{L/2}}\cM$.\\
In particular, keeping the system reflectionless (i.e., requiring relation (\ref{general reflectionless}) to be satisfied), we can make it strictly transparent from the right (left) by setting $d=\kappa$ ($d=1/\kappa$). The system is then invisible to a source (detector) placed to the system's reflectionless side.\\
For the coherent perfect absorption (CPA), employing the condition ($\cM_{11}=0$) results in the following equation:
\be
i(K_{ww}-K_{vv})=-\dfrac{\omega\kappa_{+}(\kappa-1)}{\kappa_{-}},
\ee
which is independent of the parameter $d$, and also invariant under the replacement $\kappa\rightarrow 1/\kappa$. One can simultaneously make the system transparent from the left by setting
\be
d=(\kappa^{L/2}+\kappa^{-L/2})^{2/L},
\ee
while it becomes transparent from the right if
\be
d=(\kappa^{L/2}+\kappa^{-L/2})^{-2/L}.
\ee
These results, we find in this section, have been spelled out in \cite{ghaemi-dizicheh2021compatibility} for a one-dimensional nonreciprocal lattice, where the parameter $d$ quantifies the amount of nonreciprocity and $\kappa$ captures the topological characteristics in the SSH version. We present here a general version of the compatibility and independency of transport effects for all those models categorized in the class of $r=1$. Hence, we can re-express the transport behavior in more general terms such that (1) the direction of reflectionless transport depends on the parameter $\kappa$, (2) invisibility coincides with the phase transition corresponding to $d=\kappa$ ($d=1/k$), where $d$ measures the nonreciprocity of the system, and (3) the coherent perfect absorption is compatible with transparent effect.
\subsection{Creutz Ladder Model}
A general form of the nonreciprocal one-dimensional lattice is studied in \cite{ghaemi-dizicheh2021compatibility} described by
\be
E\psi_n=V_n\psi_n+u_n\psi_{n-1}+v_{n}\psi_{n+1},
\ee
where $ V_n $ are on-site potentials, $ u_n $ are nearest-neighbor couplings from left to right, and $ v_n $ are nearest-neighbor couplings from right to left. In our formalism and notations, the hopping $\bJ_{L,R}$ and on-site $\bK$ matrices are given by
\bea
&&	\bJ_{L}=\left( \begin{array}{cc}
	0	&0  \\
	v_n& 0
\end{array}\right),~\bJ_{R}=\left( \begin{array}{cc}
	u_{n-1}	&0  \\
	0& 0
\end{array}\right),~\nonumber\\
\nonumber\\
&&~~~~~~~~~~~~\bK=\left( \begin{array}{cc}
	V_{n-1}	&u_{n}  \\
	v_{n-1}& V_n
\end{array}\right),
\eea
which leads to the same transfer matrix expressed in relation (3) of Ref. \cite{ghaemi-dizicheh2021compatibility}. We comprehensively explored the transport effect of this model in \cite{ghaemi-dizicheh2021compatibility}.\\
\begin{figure}[t]
	\includegraphics[width=\columnwidth]{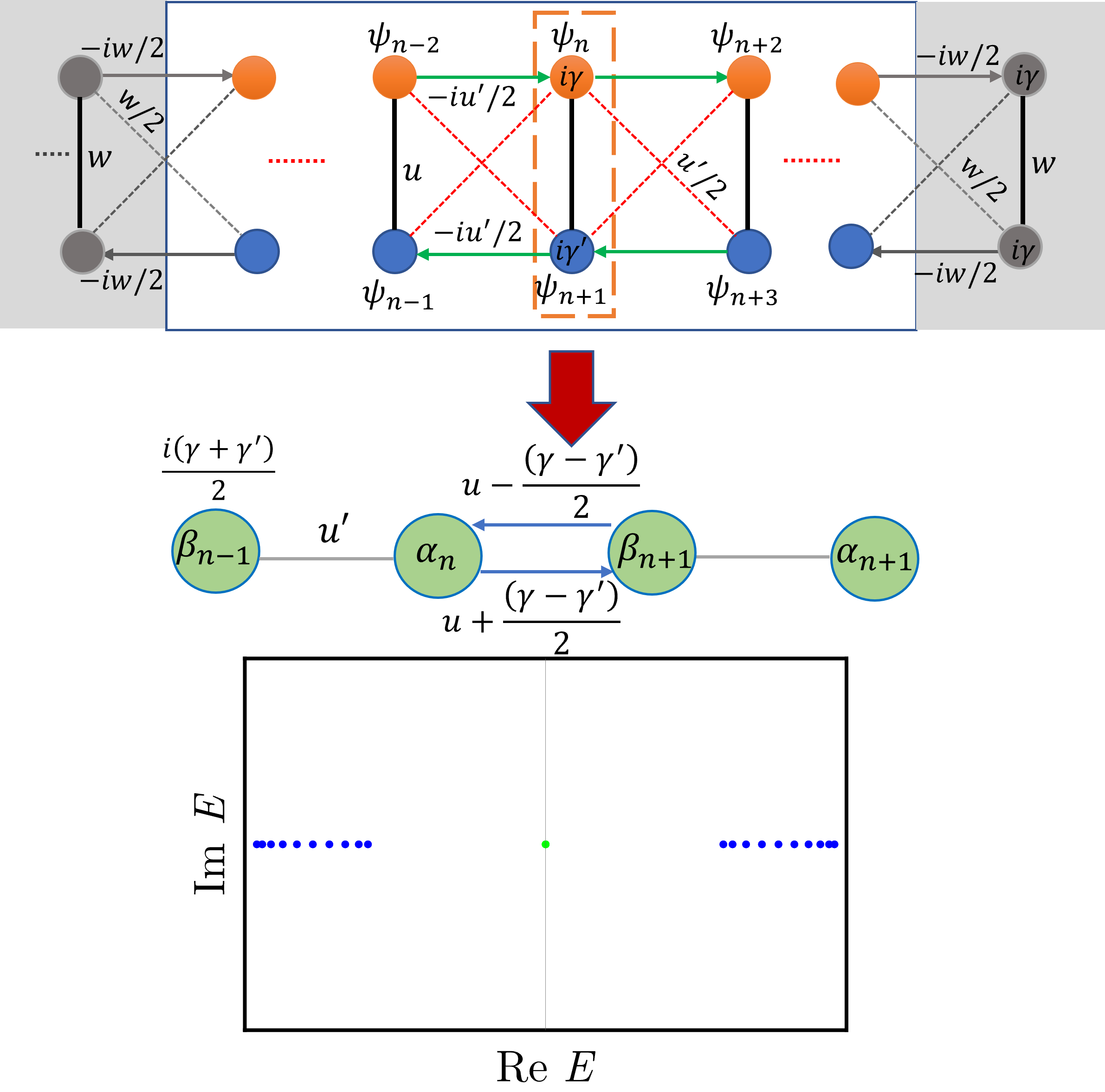}
	\caption{The Creutz ladder model of recursion relations (\ref{reclad1}) and (\ref{reclad2}), which is connected to featureless leads from both sides. The lead's structure in the diagram can be achieved by following the relation (\ref{transfromlad}) that transforms the Creutz ladder model to the non-Hermitian SSH model. This transformation reveals that the featureless leads ballistically attach to the Cruetz model with the same structure by setting $u$=$u'$=$w$. In the case of balanced gain and loss (i.e., $\gamma=\gamma'$), the system is in the non-trivial topological phase if $\vert u \vert< \vert w\vert$ and it supports a pair of degenerate zero edge modes shown by a green point in the complex energy spectrum of a finite ladder system with closed boundary conditions.} \label{fig2}
\end{figure}
In this paper, we investigate the non-Hermitian one-dimensional Creutz ladder model depicted in Fig. \ref{fig2}. This model was introduced in \cite{Yoa2018Ladder} and has been studied in other subsequent papers \cite{Kunst2018ladder,Alvarez2018ladder,liang2022topological}. Recently, a novel boundary-induced dynamical phenomenon, dubbed ``edge burst", has been observed in the ladder lattice that possesses pure gain \cite{Xue2022edgeburst}, which makes this model an interesting platform to be explored in both classical and quantum regimes.\\A tight-binding lattice describes the Creutz ladder model with the following recursion relations
\bea
&E\psi_n=i\gamma\psi_n+u\psi_{n+1}+\frac{u'}{2}(\psi_{n-1}+\psi_{n+3})&\nonumber\\
&+\frac{iu'}{2}(\psi_{n-2}-\psi_{n+2}),\label{reclad1}&\\
&E\psi_{n+1}=i\gamma'\psi_{n+1}+u\psi_{n}-\frac{iu'}{2}(\psi_{n-1}-\psi_{n+3})&\nonumber\\
&+\frac{u'}{2}(\psi_{n-2}+\psi_{n+2}).\label{reclad2}&
\eea
The coupling and on-site matrices are then given by
\be
\bJ_{L}=\left( \begin{array}{cc}
	\frac{iu'}{2}	&\frac{u'}{2}  \\
	\frac{u'}{2}& \frac{-iu'}{2}
\end{array}\right),~\bK=\left( \begin{array}{cc}
	i\gamma	&u  \\
	u& i\gamma'
\end{array}\right),
\ee
where one can see $ r=1 $. Expanding in terms of orthonormal basis by using relation (\ref{expand superstate}) gives
\be
\Psi_{n}=\left(\begin{array}{c}
	\psi_{2n-1}    \\
	\psi_{2n}   
\end{array}\right)=\frac{1}{\sqrt{2}}\left(\begin{array}{c}
	1    \\
	i   
\end{array}\right)\alpha_{n}+\frac{1}{\sqrt{2}}\left(\begin{array}{c}
	i    \\
	1   
\end{array}\right)\beta_{n}.\label{transfromlad}
\ee
It transforms the recursions relations (\ref{redladder1}) and (\ref{redladder2}) to
\bea
&E\beta_{n}=\frac{i(\gamma+\gamma')}{2}\beta_n+w\alpha_{n-1}+[u+\frac{(\gamma-\gamma')}{2}]\alpha_{n},&\label{redladder1}\\
&E\alpha_{n}=\frac{i(\gamma+\gamma')}{2}\alpha_n+w\beta_{n-1}+[u+\frac{(\gamma-\gamma')}{2}]\beta_{n},&\label{redladder2}
\eea
which can also be achieved by making use of the general recursion formulas given in (\ref{rec1}) and (\ref{rec2}). Interestingly, the above equations describe a one-dimensional SSH model in Fig. \ref{fig2}. This unitary transformation has been discussed in other literature \cite{Longi2019probingskin}. In fact, the reduction formalism leading to equations (\ref{redladder1}) and (\ref{redladder2}) reveals that the transport properties of a non-Hermitian ladder model are given by those in a generic one-dimensional tight-binding lattice studied in \cite{ghaemi-dizicheh2021compatibility} with the following on-site potentials
\be
V_{2n-1}=V_{2n}=\frac{i(\gamma+\gamma')}{2},
\ee
and couplings
\be
u_{2n}=v_{2n-1}=u',~~v_{2n}=u-\frac{(\gamma-\gamma')}{2}, ~~u_{2n}=u+\frac{(\gamma-\gamma')}{2}.
\ee
In the following, we explore the transport effects and their compatibility in detail, but before that, we make a comment on the featureless leads attached to the ladder system.\\
At the beginning of this section, we remarked on the characteristics of the attached leads in a general formalism. This will give us a clue about the lead's structure. We consider a lead with a tight-binding structure similar to the ladder model whose parameter can be read from the projected one-dimensional SSH model. It then turns out that to make the featureless lead attached ballistically to the system, we need to set $ E=i\gamma=i\gamma' $ and $ u=u'=w $ for the lead's sites. This is illustrated in Fig. \ref{fig2}.\\ For this system, the transfer matrix of the unit-cell (\ref{Transfer matrix of supercells for r1}) is 
\begin{equation}
	M_c=\sqrt{\frac{d}{\kappa}}\left( \begin{array}{cc}
		\dfrac{\left( E-i\bar{\gamma}\right)^2 }{w^2}-\kappa&\dfrac{-(E-i\bar{\gamma})}{w}  \\
		\\
		\dfrac{E-i\bar{\gamma}}{w}&-1
	\end{array}\right),\label{Transfer matrix of Ladder model}
\end{equation}
where $ \bar{\gamma}:=(\gamma+\gamma')/2 $, and $ d $ and $ \kappa $ are defined as
\begin{align}
	&d=\dfrac{2u-(\gamma-\gamma')}{2u+(\gamma-\gamma')},&&\kappa=\dfrac{4u^2-(\gamma-\gamma')^2}{4\cW^2}.
\end{align}
One can see that for $ \gamma=\gamma' $, the system has a unimodular transfer matrix. In this case, the system becomes reciprocal, and the non-Hermitian skin effect vanishes.\\
To study the transport properties of this model, we start with reflectionless transport. The corresponding boundary conditions are expressed by equation (\ref{reflectionless BC}). As we showed, the global mechanism is independent of transforming $ \kappa $ to $ 1/\kappa $, which connects the regions ($ \kappa<1 $) and ($ \kappa>1 $) in the ladder model. One then recalls that the system is topologically non-trivial with a pair of degenerate in-gap topological edge modes when \cite{Yoa2018Ladder}
\be
\sqrt{\left|  u^2-\dfrac{(\gamma-\gamma')^2}{4}\right| }<\vert w\vert,
\ee
which become zero modes for $\gamma=-\gamma'$, or equivalently $\bar{\gamma}=0$. This shows that the phase ($ \kappa<1 $) is a non-trivial topological phase where edge states do exist. Therefore, we realize that the global mechanism is essentially independent of the existence of edge states. This is analogous to what we found in \cite{ghaemi-dizicheh2021compatibility} for SSH model.\\
Now, let us turn to the local mechanism
\be
0=\cF=(E-i\bar{\gamma})^2+w^2(1-\kappa).
\ee
Interestingly, the local mechanism yields the same relation for the left and right reflectionless conditions. If we set energy to the on-site potential, i.e., $E=\bar{E}=i\bar{\gamma}$, we then find the system becomes reflectionless from both sides if $\kappa=1$, corresponding to topological phase transition
\be
\sqrt{\left|  u^2-\dfrac{(\gamma-\gamma')^2}{4}\right| }=\vert w\vert.\label{reflectioncondlad}
\ee
Also, analogously to our findings in \cite{ghaemi-dizicheh2021compatibility}, the parameter $d$ is absent in condition (\ref{reflectioncondlad}), which implies that the reflectionless transport is independent of nonreciprocity. In other words, the ladder model is reflectionless in both directions when degenerate edge states relocalize via skin effect to any edge of the system.\\ 
However, this termination might be different for other transport effects. To see this, we rewrite the total transfer matrix $\cM$ in the following form
\begin{equation}
	\cM=\frac{(-1)^{L}d^{L/2}}{2\kappa^{L/2}}\left( \begin{array}{cc}
		\kappa^{L}+1&-i(\kappa^{L}-1)  \\
		\\
		i(\kappa^{L}-1)&\kappa^{L}+1
	\end{array}\right).\label{Transfer matrix of Ladder model for pinned energy}
\end{equation}
Let us assume a system in the topological phase transition that is also reflectionless from both sides. Now, one can make it transparent from the right (left) by setting $d=\kappa$ ($d=1/\kappa$). This condition also corresponds to $\gamma-\gamma'=\fu$ ($\gamma-\gamma'=-\fu$) with $\fu=\vert u-w\vert$ showing that the invisibility direction can be changed by interchanging $\gamma\leftrightarrow\gamma'$. This coincides precisely with the skin effect phase transition of a topological state. For coherent perfect absorption (CPA), applying the condition $\cM_{11}=0$ yields $\kappa=(-1)^{1/L}$.\\
\begin{figure}[h]
	\includegraphics[scale=0.65]{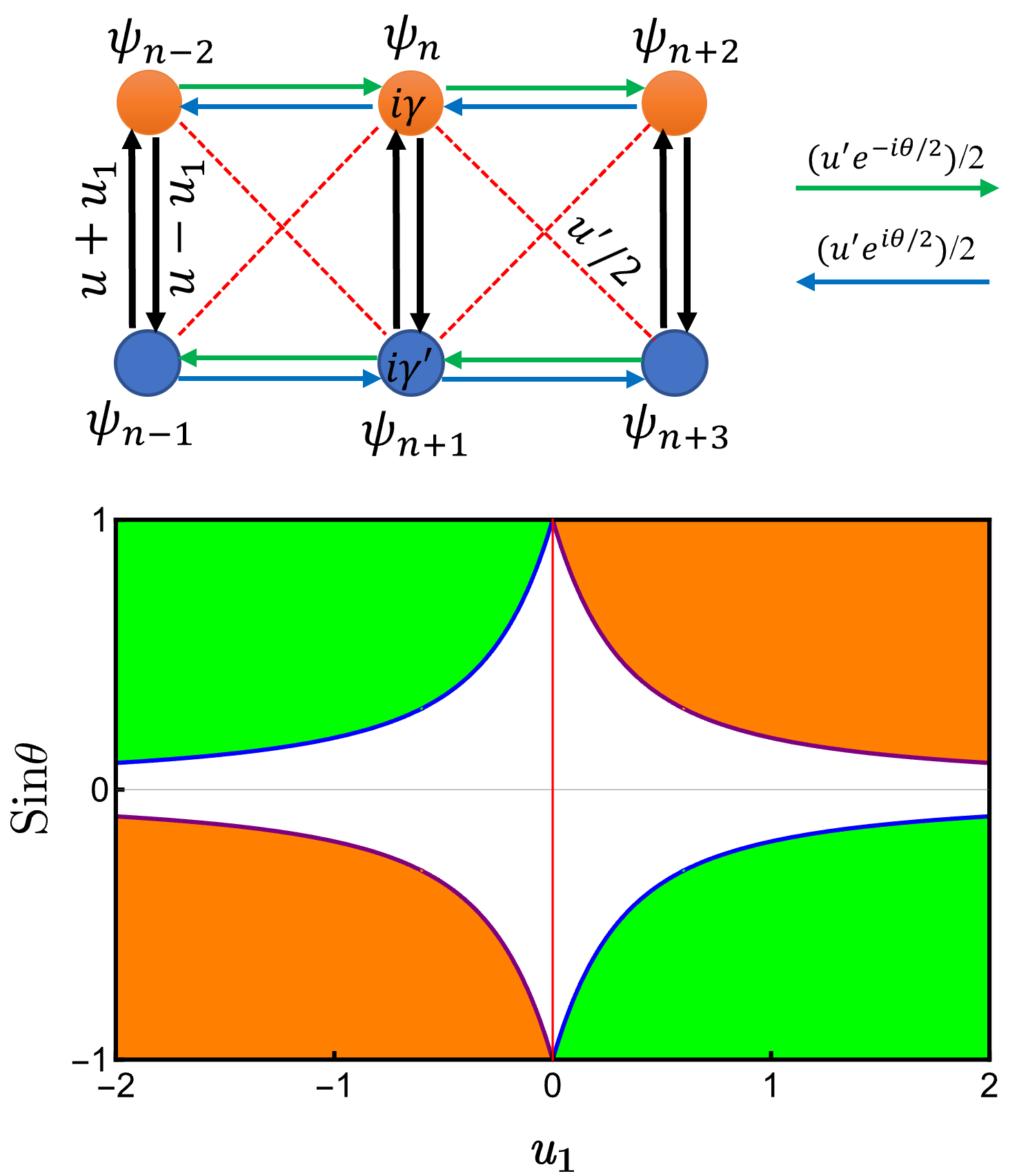}
	\caption{The possible range for parameter $\theta=\arcsin x$, where the finite system displays left or right reflectionless boundary conditions in the non-trivial topological phase for the extended Creutz Ladder illustrated schematically in the upper diagram. The blue curves locate the roots $x^{l}_{1,2}$ of Eq. (\ref{reflectionless eq}) for the left reflectionless while purple ones define the roots $x^{r}_{1,2}$ for the right reflectionless. The green and orange regions present a possible range for $\theta$ in terms of the nonreciprocity parameter $u_1$ where the system behaves reflectionless for the probe located on the right and left side, respectively. This plot shows that for a given $u_1\in[-2,2]/\{0\}$, one can find a possible range of $\theta$ that satisfies the local reflectionless condition. Those points along $u_1=0$ are excluded since the system can not satisfy the reflectionless boundary conditions in the non-trivial topological phase.} \label{fig3}
\end{figure}
The transport effects can be tuned by manipulating the parameters of the model. We show this by considering an extended version of the Creutz ladder model, with coupling $ u'e^ {-i\theta}/2$ and $ u'e^ {i\theta}/2$ between adjacent sites in the different unit-cell and $u\pm u_1$ between sites in a unit-cell \cite{liang2022topological} (see Fig. \ref{fig3}). In this model, for the lattice with pure gain or loss (i.e., $\gamma=\gamma'$), the transfer matrix is unimodular ($d=1$), while for the $\cP\cT$-symmetric system (i.e., $\gamma=-\gamma'$), one can find
\be
d=\frac{u-\gamma\sin\theta}{u+\gamma\sin\theta}.
\ee
If $\vert d\vert= 1$, there is no skin effect meaning that the ``bulk states" decay into the bulk. These states are localized on the left
boundary via non-Hermitian skin effect for $\vert d\vert< 1$ which results in $2\gamma\sin\theta<0$. For $\gamma>0$, this corresponding to $\theta\in[\pi,2\pi]$. For localizing on the right boundary, the determinant needs to satisfy $\vert d\vert>1$, leading to $2\gamma\sin\theta>0$, which includes $\theta\in[0,\pi]$. For the reflectionless transport, applying local mechanism $\cF=0$ in Eq. (\ref{reflectionless for r1}) gives
\bea
&&w^2(1-\kappa)\pm2 w u_{1}\sin\theta+\nonumber\\\nonumber
&&~~~~~\left( u\cos\theta-E-\frac{i(\gamma+\gamma')}{2}\right) ^2+u_{1}^{2}\sin^{2}\theta =0.\\
\eea
For a better illustration, let us again set $E=\bar{E}$. Then we find that the system becomes reflectionless when
\be
w^2(1-\kappa)=u^{2}(x^{2}-1)\mp2w u_{1}x-u_{1}^{2}x^{2},\label{reflectionless eq}
\ee
where $x:=\sin\theta$. For the reciprocal intercell coupling, i.e., $u_{1}=0$, the local mechanism gives
\be
w^{2}(1-\kappa)=u^{2}(x^2-1).
\ee
The left-hand side is positive in the non-trivial topological phase (i.e., $\kappa<1$). This requires $x^{2}>1$, implying that no solution exists for $\theta$. Therefore, the system can only be made reflectionless in the trivial topological phase.\\
In the case of an extra degree of freedom $u_{1}\neq0$, the right-hand side of Eq. (\ref{reflectionless eq}) is a quadratic equation whose behavior is determined by its discriminant 
\be
\Delta=4u_{1}^{2}(w^{2}-u^{2})+4u^{4}.
\ee
If $\Delta>0$, then there are two distinct real-valued roots $x_{1}$ and $x_{2}$, and the right-hand side of Eq. (\ref{reflectionless eq}) admits positive values for $x>x_{2}$ and $x<x_{1}$ where we supposed $x_{1}<x_{2}$. For $-1<x_1$ and $x_{2}<1$, one can find a possible range for $\theta$ such that the system satisfies the reflectionless boundary conditions.
In Fig. \ref{fig3}, we plot a range of possible $\theta$ in which the system is reflectionless for the incoming wave from the left (right) in its non-trivial topological phase.
	\section{Conclusions}
To summarize, we introduce a general formalism to establish transport effects and their interplays in a wide range of non-Hermitian, nonreciprocal, and potentially topological systems from a unifying scattering perspective. This work extends our previous work \cite{ghaemi-dizicheh2021compatibility} to a general formalism applicable in concrete platforms with internal degrees of freedom or in higher dimensions.\\
For specific models, we reach a global characterization of transport boundary conditions presented in table \ref{table1}, where we can retrieve our findings for the compatibility of transport properties up to a system's parameters. The model's transport behavior can be predicted based on how those parameters are interpreted in different systems. The practical realization of these systems can be investigated in nonreciprocal photonics structures such as coupled resonant optical waveguides \cite{Zhao2018microlasers,Zhu2018optica} where asymmetric coupling can be achieved by inserting an optical gain and loss medium into the link ring \cite{Zhu2020PRResearch}. Furthermore, the method described in this paper can be used to tune the parameters of the topological system in order to obtain a specific transport effect. We demonstrated this by considering a generalized version of the ladder lattice in which one of the parameters can be set to a specific range to make it reflectionless in a non-trivial topological phase. A perspective of this approach can be studied in a two-dimensional lattice such as a non-Hermitian Chern insulator, a non-Hermitian 2D Dirac semimetal, or a non-Hermitian Hofstadter model. Also, a nonlinear extension version of the transfer matrix can be considered for studying the transport effect in the nonlinear SSH model \cite{Mattis2020PRApplied}\\
	\begin{acknowledgments}
	The work on this article was initiated and partially developed during my research position at Lancaster University, and I acknowledge support from EPSRC via Programme Grant No. EP/N031776/1 in that time.\\
The author appreciates Prof. Henning Schomerus for coming up with the original idea in \cite{ghaemi-dizicheh2021compatibility} and providing valuable comments and suggestions on this paper, and Dr. Delaram Mirfendereski for reading and editing the primary draft. 
\end{acknowledgments}
\appendix
\setcounter{secnumdepth}{0}
\section{APPENDIX: TRANSFER MATRIX IN $r=2$ CLASS}\label{appendix}
\begin{figure}[t]
	\includegraphics[width=\columnwidth]{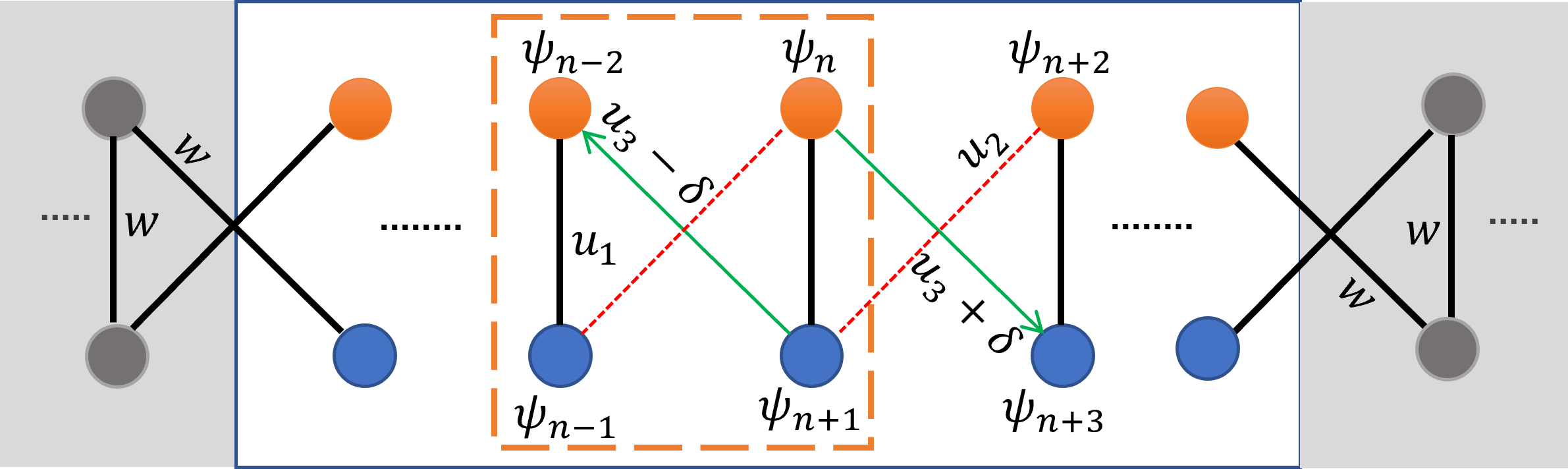}
	\caption{A SSH model with alternating $u_1$, $u_2$ couplings and a non-Hermitian $u_3\pm\frac{\delta}{2}$ coupling fitting in $r=2$ class.} \label{fig4}
\end{figure}
In this appendix, we study the transfer matrix of a specific non-Hermitian Su-Schrieffer-Heeger (SSH) model whose rank of its hopping matrix is $r=2$. The Bloch Hamiltonian of this model which is depicted in \ref{fig4} reads \cite{Song2019PRL}
\be
H(k)=d_x(k)\sigma_x+d_y(k)\sigma_y,\label{BlochH}
\ee
where $\sigma_{x,y}$ are Pauli matrices and
\bea
&&d_x(k)=u_1+(u_2+u_3)\cos k+i\frac{\delta}{2}\sin k,\\
&&d_y(k)=(u_2-u_3)\sin k+i\frac{\delta}{2}\cos k.
\eea
One can read the hopping ($J_{L,R}$) and on-site ($K$) matrices as coefficients of $e^{ik}$ and 1 in the Bloch Hamiltonian \ref{BlochH}, so that
\bea
&&	\bJ_{L}=\left( \begin{array}{cc}
	0	&u_3+\delta  \\
	u_2& 0
\end{array}\right),~\bJ_{R}=\left( \begin{array}{cc}
	0	&u_{2}  \\
	u_{3}-\delta& 0
\end{array}\right),~\nonumber\\
\nonumber\\
&&~~~~~~~~~~~~\bK=\left( \begin{array}{cc}
	0	&u_{1}  \\
	u_{1}& 0
\end{array}\right).
\eea
The SVD results in \ref{SVD} for $J_{L,R}$ with 
\bea
&&	V=\left( \begin{array}{cc}
	0	&1  \\
	1& 0
\end{array}\right),~~~~~~~~~~~~W=\left( \begin{array}{cc}
	1	&0  \\
	0& 1
\end{array}\right),~\nonumber\\
\nonumber\\
&& X_{L}=\left( \begin{array}{cc}
	u_{2}	& 0  \\
	0 & u_{3}+\delta
\end{array}\right),~X_{R}=\left( \begin{array}{cc}
u_2	&0  \\
0& u_{3}-\delta
\end{array}\right).
\eea
It meets the requirement of our formalism in which the hopping matrices $\bJ_L$ and $\bJ_R$ differ only in their singular values \cite{Note1}. The decomposition leads to the following entries of the unit-cell transfer matrix \ref{Transfer matrix of supercells}
\bea
&&	 M_{11}=\left( \begin{array}{cc}
	\frac{E^{2}}{u_{1}u_{2}}	& \frac{-2E}{u_{2}}  \\
	\frac{-2E}{u_{3}+\delta} & \frac{E^{2}}{u_{1}(u_{3}+\delta)}
\end{array}\right),~M_{12}=\left( \begin{array}{cc}
\frac{-E}{u_{1}}	& -\frac{u_{3}-\delta}{u_{2}}  \\
\frac{u_{2}}{u_{3}+\delta} & \frac{-(u_{3}-\delta)E}{u_{1}(u_{3}+\delta)}
\end{array}\right),~\nonumber\\
\nonumber\\
&& M_{21}=\left( \begin{array}{cc}
	\frac{E}{u_{1}}	& -1  \\
	-1 & \frac{E}{u_{1}}
\end{array}\right),~~~~~~~~~~M_{22}=\left( \begin{array}{cc}
-\frac{u_{2}}{u_{1}}	& 0  \\
0 & \frac{-u_{3}+\delta}{u_{1}(u_{1})}
\end{array}\right).\nonumber\\
\eea
The determinant of the transfer matrix is 
\be
d=\dfrac{u_3-\delta}{u_3-\delta},
\ee
where the transfer matrix is unimodular for the vannishing nonreciprocal parameter $\delta$ and the non-Hermitian skin effect vanishes \cite{Longi2019probingskin}. To find the reflection and transmission matrices of the system in the propagating-state basis \ref{def.propagating-space transfer matrix}, one needs to attach the entire system to the leads from both sides. We demonstrated the structure of the leads in Fig. \ref{fig4} using the formalism presented in this paper.

\end{document}